\documentclass[final,aps,prb,preprint,superscriptaddress,showpacs]{revtex4-1}

\usepackage{graphicx}
\usepackage{dcolumn}
\usepackage{amsmath,bm}
\usepackage{color}%
\usepackage{multirow}
\usepackage{array} 
\usepackage{esvect}
\usepackage{braket}
\usepackage{makecell}
\usepackage[utf8]{inputenc}
\usepackage{kotex}
\usepackage{xr}
\usepackage{soul,xcolor}
\setstcolor{red} 
\usepackage{array}

\usepackage{soul}   
\usepackage{xcolor}

\setstcolor{red}

\newif\ifclean
\cleanfalse 

\usepackage{xr}

\makeatletter
\newcommand*{\addFileDependency}[1]{
  \typeout{(#1)}
  \@addtofilelist{#1}
  \IfFileExists{#1}{}{\typeout{No file #1.}}
}
\makeatother

\newcommand*{\myexternaldocument}[1]{%
    \externaldocument{#1}%
    \addFileDependency{#1.tex}%
    \addFileDependency{#1.aux}%
}

\myexternaldocument{SI}  

\begin{document}

\title{Propagation-mediated amplification of \{11\={2}0\}-biased inversion domain boundary alignment in polarity-mixed GaN lateral overgrowth}

\author{Harim Song}
\affiliation{Department of Physics and Research Institute for Basic Sciences, Kyung Hee University, Seoul, 02447, Republic of Korea}

\author{Donghoi Kim}
\affiliation{Department of Information Display, Kyung Hee University, Seoul, 02447, Republic of Korea}

\author{Chinkyo Kim}
\email[Corresponding author. E-mail:]{ckim@khu.ac.kr}
\affiliation{Department of Physics and Research Institute for Basic Sciences, Kyung Hee University, Seoul, 02447, Republic of Korea}
\affiliation{Department of Information Display, Kyung Hee University, Seoul, 02447, Republic of Korea}


\begin{abstract}
GaN polarity inversion and the associated inversion domain boundaries (IDBs) are frequently observed during lateral overgrowth and are often discussed in terms of the small energetic spread among competing IDB structures predicted by first-principles calculations. In circular mask openings, \(\{11\bar{2}0\}\)-aligned IDBs have previously been explained by geometric closure of a single-polarity hexagonal domain at the circular boundary. Here we examine an experimentally distinct regime in which opposite-polarity domains already coexist within the opening before the later development of long, straight IDB traces. In this mixed-polarity regime, the final trace orientation cannot be attributed solely to the macroscopic circular boundary. Nevertheless, plan-view SEM line-trace statistics show that IDB orientations remain biased toward the \(\{11\bar{2}0\}\) family. To quantify how this bias develops during propagation, we perform distance-resolved, length-weighted orientation analysis in concentric annular regions defined from the opening center. The resulting metrics show that \(\{11\bar{2}0\}\)-biased alignment is progressively amplified with propagation distance, while the orientation distribution becomes narrower, indicating systematic sharpening of the preferred alignment state. We further apply the same ring-resolved statistical operators to minimal two-domain propagation simulations in a circular opening and find that a propagation-mediated anisotropy reproduces the observed radial amplification under fixed circular geometry. Together, these results establish a quantitative phenomenology of \(\{11\bar{2}0\}\)-biased IDB alignment in polarity-mixed GaN lateral overgrowth on patterned sapphire and indicate that, although mask-boundary-imposed selection may describe single-polarity closure cases, the present mixed-polarity regime is better explained by propagation-mediated amplification.\end{abstract}
\maketitle

\section{Introduction}
Wurtzite GaN is a polar semiconductor in which crystallographic polarity (Ga-polar [0001] vs.\ N-polar [000\={1}]) strongly influences surface chemistry, adatom kinetics, and defect formation during epitaxy.\cite{Hellman-IJNSR-3-11} Consequently, controlling and diagnosing polarity remain central issues in GaN growth on foreign substrates and patterned templates, where local polarity inversion or mixed-polarity regions can emerge depending on nucleation conditions, impurities, and growth mode.\cite{Sumiya-IJNSR-9-1} A comprehensive review by Z\'u\~niga-P\'erez \emph{et al.} summarizes polarity-related theory, measurement methodologies, and growth-dependent polarity evolution in GaN and related wurtzite materials, emphasizing that polarity is often spatially heterogeneous and strongly process dependent.\cite{Zuniga-APR-3-041303}

Polarity inversion is accompanied by the formation of inversion domain boundaries (IDBs), which are extended planar defects separating oppositely oriented polar domains. First-principles calculations have established representative domain-wall structures and energies for IDBs and related stacking-mismatch boundaries in GaN, providing a thermodynamic baseline for discussing which atomic-scale IDB variants may be favored.\cite{Northrup-PRL-77-103,Umar-PRB-103-165305} Because the calculated energy differences among competing IDB domain-wall structures can be small, thermodynamics alone does not necessarily impose a unique macroscopic trace selection; accordingly, one might expect no overwhelming preference in the \emph{observed plan-view IDB trace families} (e.g., \(\{1\bar{1}00\}\) versus \(\{11\bar{2}0\}\)) unless additional kinetic or geometric selection rules operate during boundary formation and evolution.\cite{Hwang-ACSAEM-6-3257}

Patterned-mask epitaxy and epitaxial lateral overgrowth (ELOG) provide a controlled platform for interrogating polarity inversion and IDB crystallography under well-defined geometric and kinetic constraints.\cite{Beaumont-JCG-189-97,Hiramatsu-JCG-221-316} In circular-patterned SiO\(_2\) mask systems, Kim \emph{et al.} reported lateral polarity inversion triggered at the mask boundary during ELOG and observed that IDB traces formed preferentially along \(\{11\bar{2}0\}\).\cite{Kim-JAC-51-1551} Lee \emph{et al.} further demonstrated that Ga-to-N polarity inversion can occur on the flat region beyond the edge of a circular-patterned SiO$_2$ mask under appropriate growth conditions, indicating that inversion pathways are not restricted to strictly edge-triggered scenarios.\cite{Lee-JAC-52-532} These studies established an important geometric baseline in which mask-boundary-triggered inversion and subsequent closure of a single-polarity domain can account for the final IDB trace morphology.

Despite these advances, an important question remains: how is IDB trace crystallography established when Ga- and N-polar domains already coexist within the growth region, i.e., when the final trace orientation can no longer be attributed solely to mask-boundary-imposed selection?  In this mixed-polarity regime, alignment may instead be governed more appropriately by propagation-mediated processes that progressively shape boundary orientation during growth, rather than by the initial inversion trigger or single-domain closure geometry alone.  Recent work showing that growth-mode control, for example through the establishment of step-flow, can strongly suppress inversion domains further highlights the potential for kinetics to dominate over small thermodynamic differences.\cite{Zhang-CGD-23-1049}

Here we address this gap by establishing a quantitative, statistics-based description of IDB trace alignment during polarity-mixed lateral overgrowth on patterned sapphire. Crucially, \emph{experimental} polarity mapping (based on etch-contrast/morphology criteria and co-registered plan-view microscopy; see Methods) confirms that Ga- and N-polar domains already coexist within circular openings \emph{prior} to the emergence of long, straight IDB traces. The present system therefore lies in a distinct mixed-polarity regime for which mask-boundary-imposed selection alone is insufficient. Within the same circular-opening geometry, we nevertheless observe that IDB orientations remain biased toward the \(\{11\bar{2}0\}\) family. Using length-weighted orientation statistics of plan-view SEM line traces and distance-resolved analysis within concentric annular regions, we further show that \(\{11\bar{2}0\}\)-biased alignment is progressively amplified with propagation distance, as quantified by the orientation-alignment parameter \(\kappa(r)\) and the concomitant narrowing of the orientation distribution. Together, these measurements establish a stringent experimental benchmark for the present mixed-polarity regime and indicate that, whereas mask-boundary-imposed selection may account for previously studied single-polarity closure cases, the observed alignment here is better described by propagation-mediated amplification. In the following, we use minimal two-domain propagation simulations, in which random nucleation of opposite-polarity domains generates a mixed-polarity configuration within a circular opening, as a quantitative testbed to examine whether an orientation-dependent propagation ingredient can reproduce the observed radial amplification within the same circular-opening geometry.

\section{Experimental and computational methods}

\subsection{Growth of GaN on SiO$_2$-patterned $c$-plane sapphire substrates}
A SiO$_2$ mask was deposited on $c$-plane sapphire and lithographically patterned to form circular openings of diameter \(\sim 4~\mu\mathrm{m}\) arranged in a hexagonal array. GaN was grown on the patterned substrates by hydride vapor phase epitaxy (HVPE). Polarity contrast was verified by KOH etching, which produces distinct etch morphologies for Ga- and N-polar surfaces.\cite{Li-JAP-90-4219} The assignment of SEM line traces to IDBs was validated by comparing the same GaN domains within individual circular openings before and after KOH etching, which revealed polarity-dependent etch morphologies across the traced boundaries.\cite{Li-JAP-90-4219,Zhuang-MSER-48-1} These etch-derived polarity assignments were used as the ground truth for the polarity-resolved IDB analysis described below.

\subsection{Polarity segmentation and ring-resolved IDB extraction}
Plan-view SEM images were processed in Fiji/ImageJ (v1.54p).\cite{Schindelin-NM-9-676} Within Fiji, polarity domains were segmented using supervised classification (Trainable Weka Segmentation), the resulting label map was converted into binary masks for the two polarity domains, and the mutual-contact boundary between the two masks was extracted and reduced to a one-pixel-wide skeleton.\cite{Arganda-BI-33-2424}  To evaluate radial evolution within a circular opening, the opening region of interest (ROI) was defined from the segmented image, and its center was taken as the ROI centroid, i.e., the geometric center of the segmented opening.  A set of concentric circular boundaries was then generated from this centroid within the opening. The center-only region was defined as the innermost disk enclosed by the first boundary, whereas Rings~1--4 were defined as successive annular regions between adjacent boundaries, such that Ring~1 denotes the innermost annulus and higher ring indices denote progressively more outward annuli closer to the opening boundary. The IDB skeleton was intersected with each region to yield ring-resolved IDB skeletons for subsequent orientation analysis.

\subsection{Segment-based orientation histograms and alignment metrics}

Ring-resolved IDB skeletons were exported as binary images and analyzed with a custom Python script. The one-pixel-wide skeleton in each region was decomposed into connected traces, which were then represented as polylines. Local orientations were estimated by principal-component analysis (PCA) of short sliding windows along each polyline.\cite{Jolliffe-PTRSA-374-20150202} In the final analysis, each PCA window contained 7 consecutive skeleton points, the window was advanced by 3 points at a time, and the resulting orientations were accumulated into a histogram with 180 bins over \(0\le\theta<180^\circ\). For each sliding window, the PCA principal axis was taken as a single representative in-plane orientation \(\theta\in[0,180^\circ)\), and the arc length of that window was used as its weight. These weighted contributions were then mapped onto the projected orientation coordinate \(x\) for quantitative analysis. The selected window size is small enough to resolve local IDB orientation while remaining large enough to suppress pixel-scale discretization noise. More importantly, the main conclusion of this work is not tied to a single arbitrary parameter choice, but to the persistence of the radial alignment trend under reasonable variation of the analysis parameters.

To quantify preferential alignment toward crystallographic direction families, we introduce the orientation-alignment parameter \(\kappa\in[-1,1]\) by mapping \(\theta\) onto a family score \(x(\theta)\), such that \(x=+1\) corresponds to perfect \(\{11\bar{2}0\}\) alignment and \(x=-1\) corresponds to perfect \(\{1\bar{1}00\}\) alignment. The alignment parameter is then computed as the length-weighted mean
\begin{equation}
\kappa=\sum_i x_i w_i,
\end{equation}
where \(w_i\) is the normalized length fraction in bin \(i\), with \(\sum_i w_i=1\). Radial evolution is reported for the center-only region and across the annular ROIs as \(\kappa(r)\). From the corresponding normalized distribution \(P(x)\), we also evaluate the distribution width \(\sigma_x\) and the branch/tail probabilities described in the Results section.

\subsection{Kinetic simulation of IDB time evolution (C implementation)}

To test whether propagation-mediated anisotropy can reproduce the experimentally observed radial amplification of \(\{11\bar{2}0\}\)-biased alignment in the mixed-polarity regime, we implemented a two-domain time-evolution simulator in C based on a level-set description of interface propagation on a 2D Cartesian lattice representing a single patterned opening.\cite{Osher-LSM} Interface motion was governed by a kinetic-Wulff-type anisotropic velocity law with alternating fast and slow direction families,\cite{Du-PRL-95-155503,Kim-JAC-51-1551} so that propagation bias could be introduced in a controlled manner under mixed-polarity conditions. The representative simulations discussed in this work were initialized by random nucleation of opposite-polarity domains within a circular mask, which gives rise to a mixed-polarity state prior to IDB evolution.

The C simulation was used only to generate the evolving mixed-polarity domain structure and the corresponding IDB geometry. The resulting simulated IDB images were then analyzed using the same Python-based PCA workflow applied to the experimental IDB skeleton images. In this way, the projected orientation distribution \(P(x)\), the orientation-alignment parameter \(\kappa\), the distribution width \(\sigma_x\), and the branch/tail probabilities were obtained using the same measurement procedure as for the experimental data. Accordingly, the simulation serves here as a minimal quantitative testbed for propagation-mediated amplification, rather than as a definitive microscopic model of IDB selection.

\section{Results and discussion}

\subsection{IDB observation in GaN grown on patterned substrates}

\begin{figure}
\centering
\includegraphics[width=0.80\columnwidth]{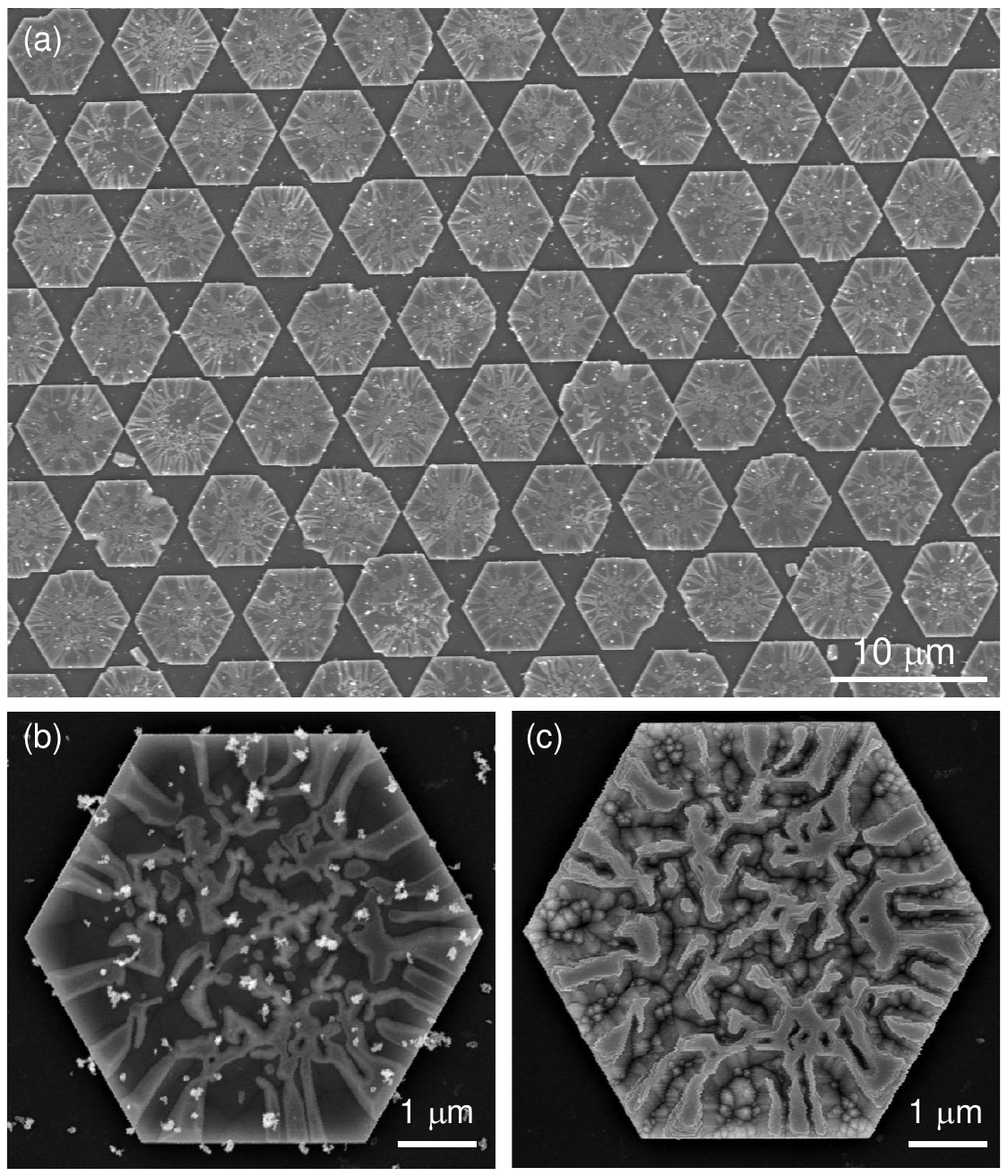}
\caption{(a) Large-area plan-view SEM image of GaN grown on a SiO$_2$-patterned sapphire substrate, showing IDB-like line traces across many circular openings. (b) Pre-KOH-etching SEM image of a representative GaN domain grown from a single circular opening. (c) Post-KOH-etching SEM image of the same GaN domain. The KOH etch reveals polarity-dependent surface morphologies, \textit{i.e.,} heavily etched N-polar GaN versus nominally inert Ga-polar GaN. Direct comparison before and after etching shows that the pre-etch line traces coincide with the boundary between differently etched Ga- and N-polar regions, thereby confirming that the analyzed traces correspond to inversion domain boundaries (IDBs). The larger-area view further indicates that the qualitative tendency toward preferentially aligned straight IDB traces is reproduced across many openings in the patterned array.}
\label{fig:IDB_SEM}
\end{figure}

Figure~\ref{fig:IDB_SEM} shows representative plan-view SEM images of GaN grown on a SiO$_2$-patterned sapphire substrate. The larger-area SEM view in Fig.~\ref{fig:IDB_SEM}(a) shows that IDB-like line traces are observed across many circular openings in the patterned array, indicating that the qualitative tendency toward preferentially aligned straight traces is reproduced over a broad area. Figures~\ref{fig:IDB_SEM}(b) and \ref{fig:IDB_SEM}(c) show, respectively, the same representative GaN domain within a single opening before and after KOH etching. In the post-etch image, the boundary between the differently etched domains coincides with the pre-etch trace, confirming that the line features observed in plan-view SEM are true inversion domain boundaries (IDBs) separating Ga- and N-polar GaN. The same polarity-dependent etch contrast further shows that opposite-polarity domains already coexist within the opening before the development of long, straight IDB segments. This point is important because it shows that the preferential appearance of \(\{11\bar{2}0\}\)-aligned IDBs in the present system cannot be attributed solely to the canonical single-domain, vertex-driven closure picture.

Instead, the observations indicate that the present system lies in a mixed-polarity regime for which mask-boundary-imposed selection alone is insufficient. Although the detailed ring-resolved analysis below is carried out for a representative opening, the larger-area SEM image indicates that the underlying qualitative behavior is not unique to that opening. The opening selected for detailed analysis was therefore taken as a representative case for quantitative distance-resolved analysis. This motivates the ring-resolved analysis presented below, including the center-only region, to quantify how the IDB orientation distribution evolves from the interior of the opening toward successive outer annuli.

\subsection{Ring-resolved orientation analysis and radial evolution}

Using the length-weighted orientation distributions obtained from the ring-resolved IDB skeletons (Methods), we construct the corresponding normalized distribution \(P(x_i)\) in the projected orientation coordinate \(x_i\in[-1,1]\), satisfying
\begin{equation}
\sum_i P(x_i)=1.
\end{equation}
In the present convention, \(x=-1\) corresponds to the \(\{1\bar{1}00\}\) orientation and \(x=+1\) corresponds to the \(\{11\bar{2}0\}\) orientation.

The primary scalar measure of alignment is the orientation-alignment parameter \(\kappa\) defined in Eq.~(1), which represents the center of mass of the distribution along the projected orientation coordinate. By construction, positive \(\kappa\) indicates preferential alignment toward the \(\{11\bar{2}0\}\) family, whereas negative \(\kappa\) indicates preferential alignment toward the \(\{1\bar{1}00\}\) family. To characterize the spread of the distribution, we use the weighted standard deviation
\begin{equation}
\sigma_x = \sqrt{\sum_i P(x_i)\,(x_i-\kappa)^2}.
\end{equation}
A smaller value of \(\sigma_x\) therefore corresponds to a narrower distribution and, in the present context, to progressive peak sharpening.

To separately evaluate the spectral weight on the positive, negative, and central portions of the orientation axis, we define
\begin{equation}
P_{+} = \sum_{x_i>0} P(x_i),
\qquad
P_{-} = \sum_{x_i<0} P(x_i),
\qquad
P_{0} = P(x=0).
\end{equation}
We also monitor the accumulation of probability near the limiting orientations through
\begin{equation}
\begin{aligned}
P(x\ge 0.8) &= \sum_{x_i\ge 0.8} P(x_i),\\
P(x\le -0.8) &= \sum_{x_i\le -0.8} P(x_i).
\end{aligned}
\end{equation}
Here, \(P_{+}\), \(P_{-}\), and \(P_{0}\) quantify the total probability residing on the \(x>0\), \(x<0\), and central portions of the orientation coordinate, respectively, whereas \(P(x\ge 0.8)\) and \(P(x\le -0.8)\) quantify the fraction of the distribution concentrated near the \(+1\) and \(-1\) extremes.

\begin{figure}
\centering
\includegraphics[width=0.65\columnwidth]{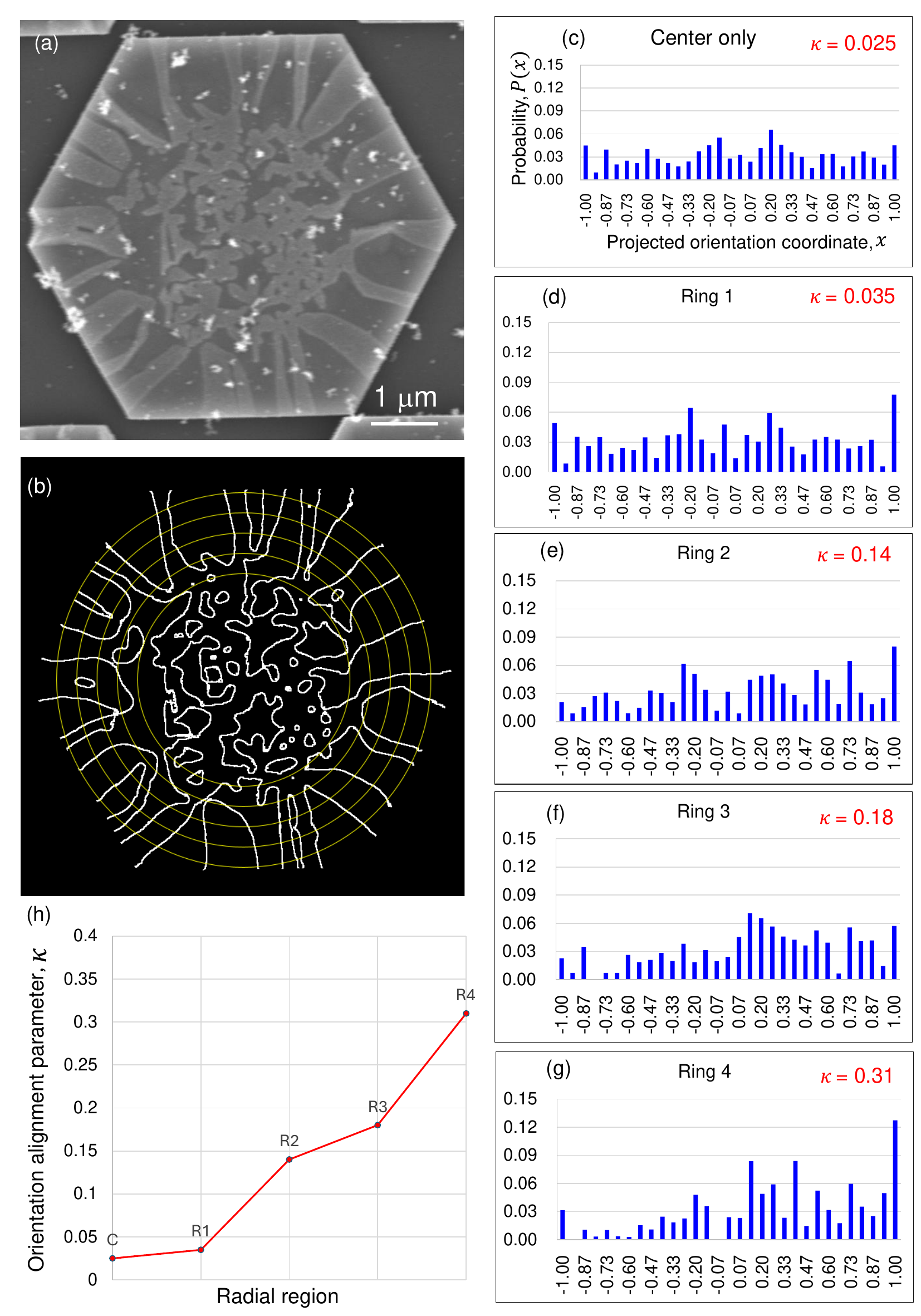}
\caption{(a) Representative SEM image of a GaN domain grown on a SiO$_2$-patterned $c$-plane sapphire substrate, shown here as the starting point for the detailed ring-resolved orientation analysis. (b) Ring-resolved IDB skeleton extracted from the SEM image in panel (a), including the center-only region and successive concentric rings. The center-only region is defined by the innermost circle, whereas Rings~1--4 correspond to the annular regions between adjacent concentric circles. (c)--(g) Normalized projected orientation distributions, \(P(x)\), for the center-only region and Rings~1--4, plotted as a function of the reduced orientation coordinate \(x\). Here, \(x=+1\) and \(x=-1\) correspond to \(\{11\bar{2}0\}\) and \(\{1\bar{1}00\}\) alignment, respectively, and higher ring indices denote larger radial distances from the opening center. The orientation-alignment parameter \(\kappa\) is shown in each panel. (h) Radial evolution of \(\kappa\). C denotes the center-only region, and R1--R4 denote the first through fourth radial rings.}
\label{fig:radial_histogram}
\end{figure}

Figure~\ref{fig:radial_histogram} shows the ring-resolved orientation distributions within a representative opening. The center-only region exhibits a nearly symmetric distribution with only a weak net bias, whereas the outer rings show progressively stronger accumulation of weight on the \(x>0\) side. In Rings~1--4, the probability near \(x=+1\), corresponding to the \(\{11\bar{2}0\}\) family, becomes increasingly prominent, while the weight near \(x=-1\), corresponding to \(\{1\bar{1}00\}\), is progressively reduced. Consistent with these distributions, Fig.~\ref{fig:radial_histogram}(h) shows that the orientation-alignment parameter \(\kappa(r)\) increases with ring index.

\begin{figure}
\centering
\includegraphics[width=0.610\columnwidth]{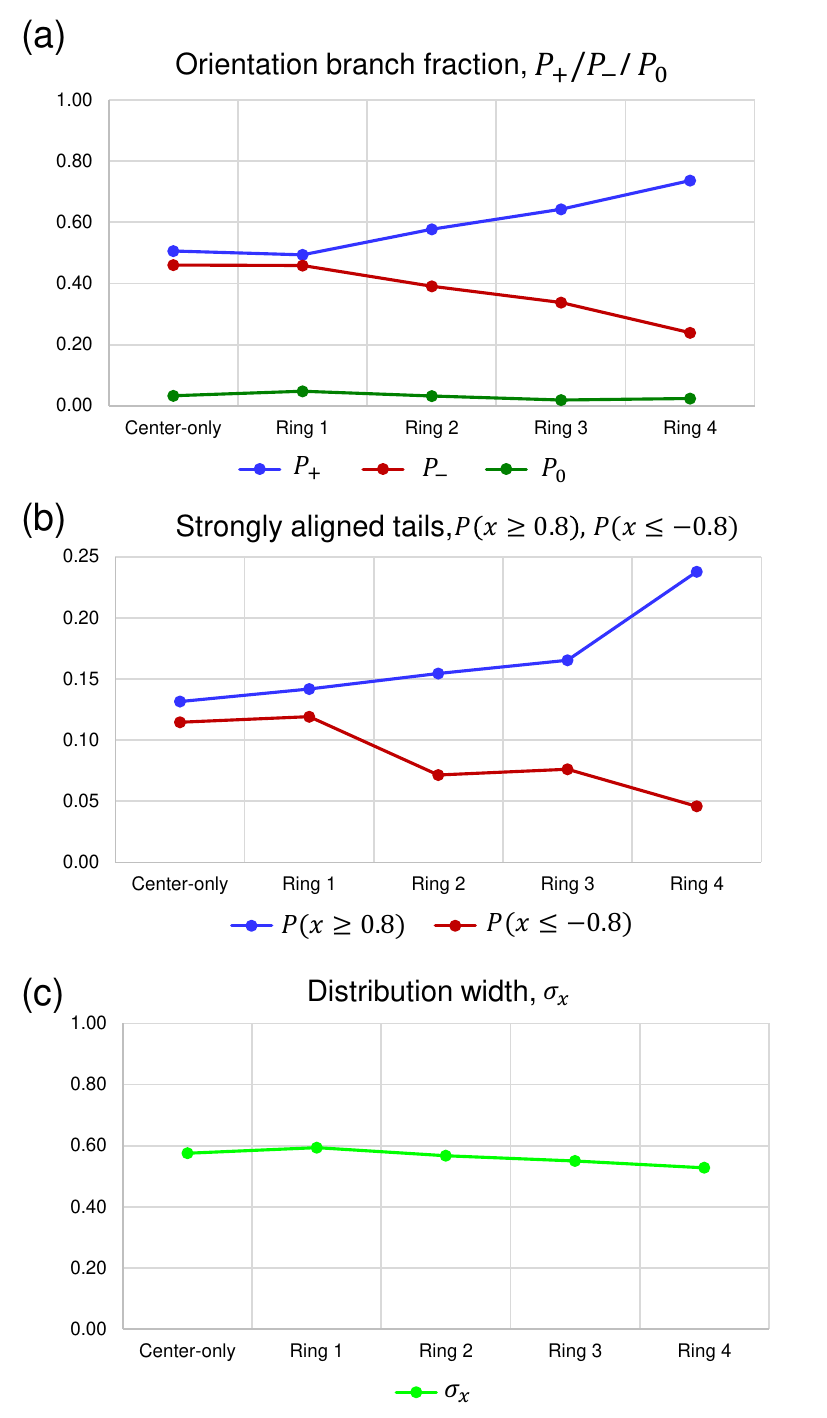}
\caption{Summary metrics for distance-resolved IDB alignment in the SEM image. (a) Orientation branch fractions \(P_{+}\), \(P_{-}\), and \(P_{0}\), where \(P_{+}\) and \(P_{-}\) denote the integrated probability on the \(x>0\) and \(x<0\) sides, corresponding to the \(\{11\bar{2}0\}\)- and \(\{1\bar{1}00\}\)-biased branches, respectively, and \(P_{0}\) denotes the weight at the central bin. (b) Strongly aligned tail fractions \(P(x \ge 0.8)\) and \(P(x \le -0.8)\), which quantify the probability concentrated near the \(\{11\bar{2}0\}\)- and \(\{1\bar{1}00\}\)-proximate limits, respectively. (c) Distribution width \(\sigma_x(r)\), where smaller \(\sigma_x\) indicates a narrower distribution and therefore stronger peak sharpening. All quantities in (a)--(c) are evaluated from length-weighted orientation histograms normalized such that \(\sum_i P(x_i)=1\) within each region.}
\label{fig:ring_metrics}
\end{figure}

The distance-resolved statistics summarized in Fig.~\ref{fig:ring_metrics} further constrain the interpretation of the radial evolution. Relative to the center-only region, the outer rings show increasing weight on the \(x>0\) side and decreasing weight on the \(x<0\) side, consistent with progressive strengthening of the \(\{11\bar{2}0\}\)-biased branch. The strongly aligned \(\{11\bar{2}0\}\)-proximate tail \(P(x\ge 0.8)\) also increases toward the outer rings, whereas the competing \(\{1\bar{1}00\}\)-proximate tail \(P(x\le -0.8)\) is overall suppressed despite a slight local rebound at Ring~3. At the same time, the distribution width decreases from \(\sigma_x=0.594\) in Ring~1 to \(0.567\), \(0.550\), and \(0.528\) in Rings~2, 3, and 4, respectively, consistent with progressive peak sharpening during boundary advance. Taken together, the branch fractions, tail probabilities, and narrowing of \(\sigma_x\) show that the \(\{11\bar{2}0\}\) bias is progressively amplified with propagation distance under an otherwise unchanged circular-opening geometry. The observations therefore indicate that the alignment is not accounted for by mask-boundary-imposed selection alone, but instead requires propagation-mediated amplification in the present mixed-polarity regime.

\subsection{Simulation test of propagation-mediated amplification}

The experimental results above motivate a more focused question: whether a minimal orientation-dependent propagation ingredient is sufficient to reproduce both the sign of the bias and its radial amplification under otherwise unchanged circular-opening geometry. To address this question, we analyze minimal two-domain propagation simulations using the same statistical operators applied to the experimental data.

To interpret the robust \(\{11\bar{2}0\}\)-biased IDB alignment observed under polarity-mixed lateral overgrowth, we implement a two-domain propagation simulation as a minimal surrogate for propagation-mediated anisotropy. The objective is not to assign a unique microscopic mechanism from morphology alone, but to test whether an orientation-dependent propagation ingredient can reproduce both the \emph{sign} of the measured bias and its systematic radial amplification when the same analysis operators used for the experiment are applied to the simulated structures. In this sense, the simulation serves as a quantitative testbed for propagation-mediated amplification rather than as a definitive microscopic model of IDB selection.

\begin{figure}
\centering
\includegraphics[width=0.77\columnwidth]{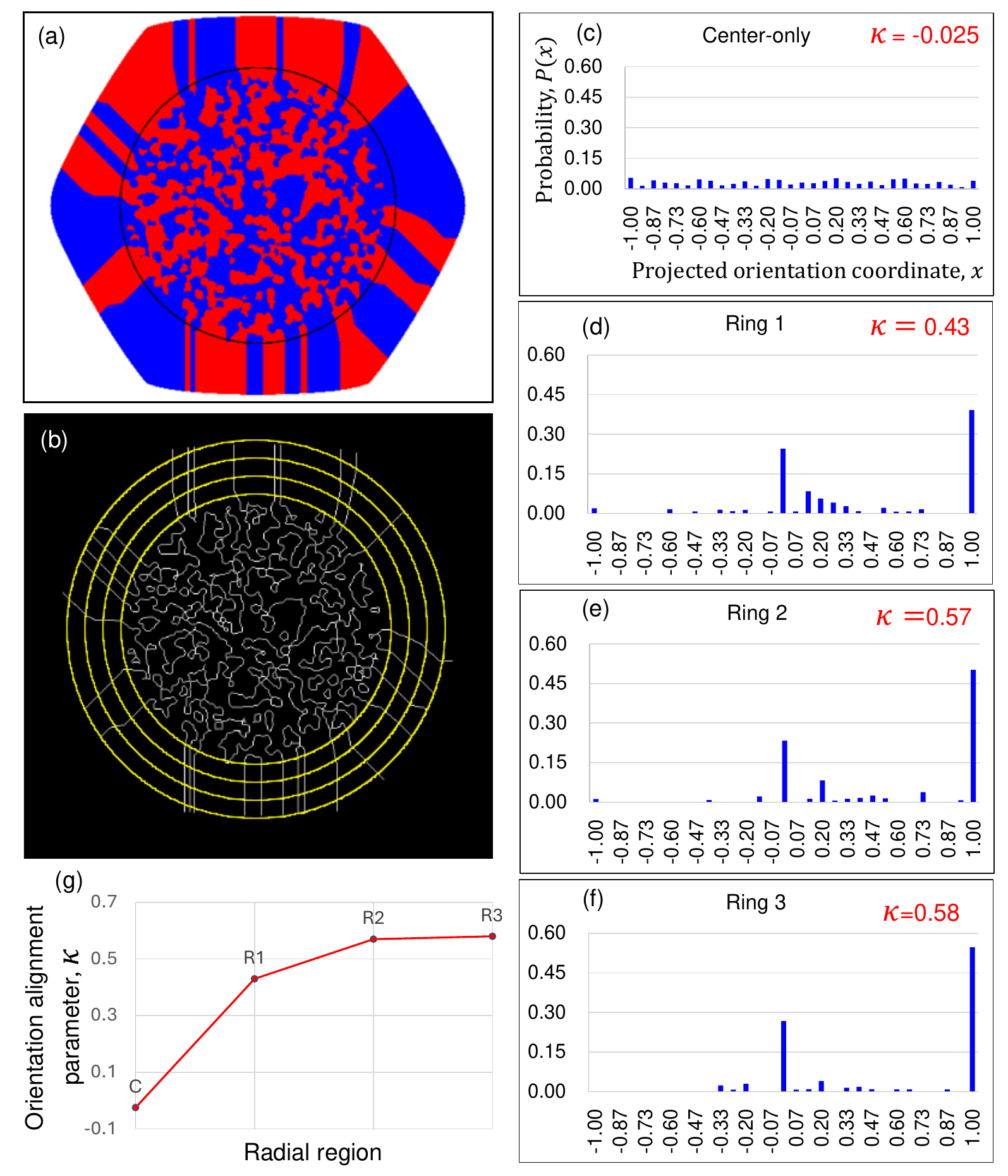}
\caption{(a) Representative simulated morphology for IDB propagation following random nucleation of opposite-polarity domains within a circular opening. (b) Ring-resolved IDB skeleton extracted from the simulated structure, including the center-only region and successive concentric rings. The center-only region is defined by the innermost circle, whereas Rings~1--3 correspond to the annular regions between adjacent concentric circles. (c)--(f) Normalized projected orientation distributions, \(P(x)\), for the center-only region and Rings~1--3, plotted as a function of the reduced orientation coordinate \(x\). Here, \(x=+1\) and \(x=-1\) correspond to \(\{11\bar{2}0\}\) and \(\{1\bar{1}00\}\) alignment, respectively, and higher ring indices denote larger radial distances from the opening center. The orientation-alignment parameter \(\kappa\) is shown in each panel. (g) Radial evolution of \(\kappa\). The same analysis procedure used for the SEM data is applied to the simulated IDB skeleton in order to compare the radial evolution of orientation alignment on an equal footing.}
\label{fig:IDB_simulation}
\end{figure}

The center-only region in the representative simulation remains nearly symmetric, with only a slight negative bias (\(\kappa=-0.025\)), and thus serves as the radial baseline for the subsequent ring-resolved evolution. The ring-resolved orientation distributions shown in Fig.~\ref{fig:IDB_simulation} then exhibit a systematic evolution toward the \(x>0\) side, corresponding to the \(\{11\bar{2}0\}\) orientation family. The orientation-alignment parameter increases from \(\kappa=0.43\) for Ring~1 to \(0.57\) and \(0.58\) for Rings~2 and 3, respectively, indicating a progressive shift of the distribution weight toward positive \(x\). In addition to the dominant peak near \(x=+1\), the simulation also exhibits a finite peak near \(x\approx 0\). This feature does not imply a distinct preferred crystallographic alignment at this value of the reduced coordinate. Rather, it reflects the presence of intermediate in-plane orientations that remain away from both limiting families and are combined in the projected \(x\)-representation. Accordingly, the bimodal-like distribution observed in the simulation, with dominant weight near \(x\approx 1\) and residual weight near \(x\approx 0\), should not be interpreted as stronger alignment than in the experiment. Instead, it reflects the simplified nature of the minimal model, in which intermediate orientations are not fully resolved, giving rise to an apparent two-state-like distribution.

\begin{figure}
\centering
\includegraphics[width=0.65\columnwidth]{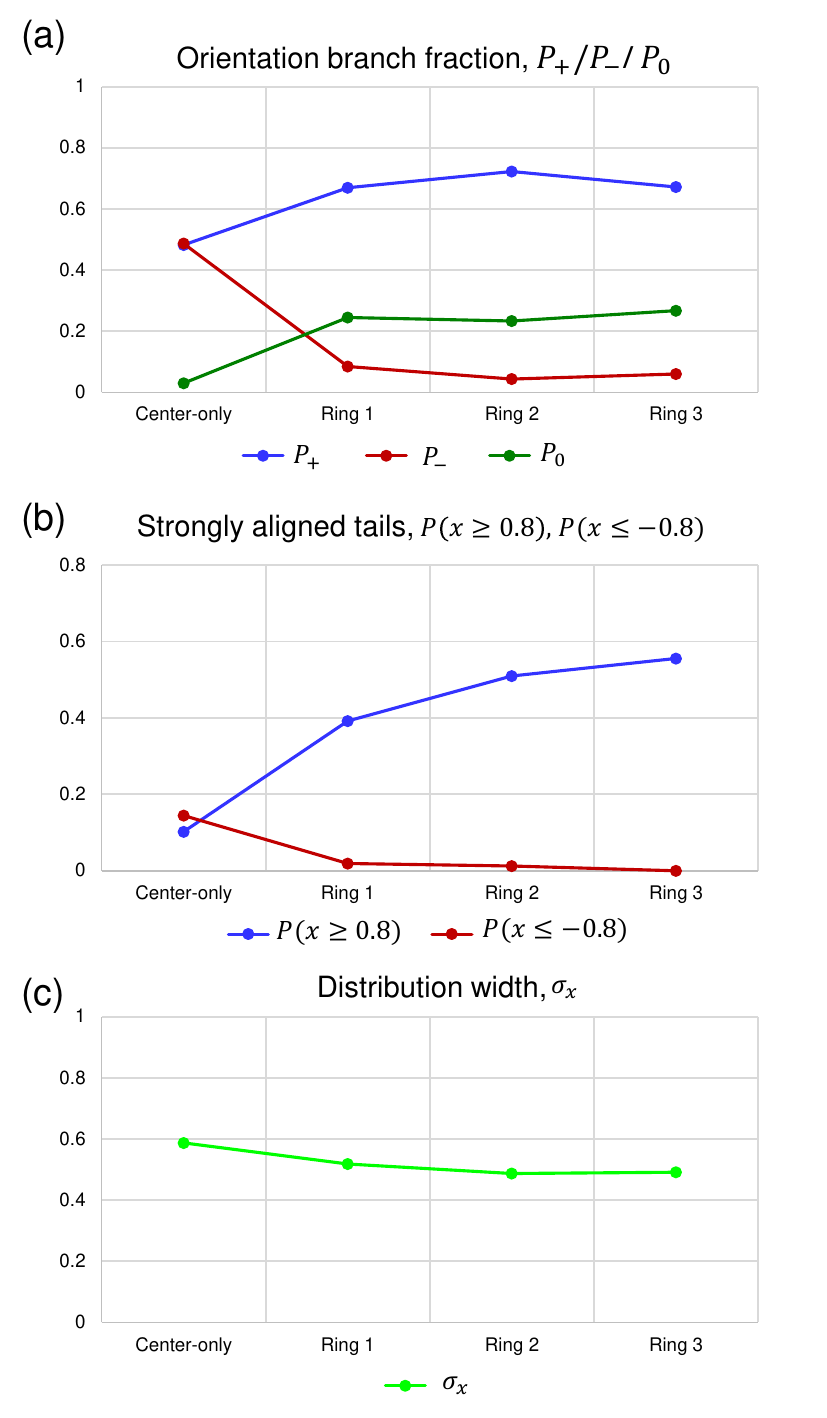}
\caption{Distance-resolved summary metrics for the simulated IDB structure (center-only and Rings~1--3), evaluated using the same statistical definitions as for the experimental data. (a) Orientation branch fractions \(P_{+}\), \(P_{-}\), and \(P_{0}\). (b) Strongly aligned tail fractions \(P(x \ge 0.8)\) and \(P(x \le -0.8)\). (c) Distribution width \(\sigma_x(r)\). All metrics in (a)--(c) are derived from length-weighted orientation histograms that are normalized independently within each region such that \(\sum_i P(x_i)=1\).}
\label{fig:simulation_ring_metrics}
\end{figure}

The distance-resolved summary metrics in Fig.~\ref{fig:simulation_ring_metrics} reinforce this interpretation. The weighted standard deviation decreases from \(\sigma_x=0.52\) in Ring~1 to \(0.49\) in Rings~2 and 3, evidencing gradual peak sharpening. Consistent with this trend, \(P_{+}\) remains much larger than \(P_{-}\) for all rings, while the strongly aligned \(\{11\bar{2}0\}\)-proximate tail \(P(x\ge 0.8)\) increases from 0.392 in Ring~1 to 0.510 and 0.556 in Rings~2 and 3, respectively. In contrast, the competing \(\{1\bar{1}00\}\)-proximate tail \(P(x\le -0.8)\) remains small and vanishes entirely in Ring~3. Taken together, these results show that the simulated distribution becomes increasingly concentrated near the \(x=+1\) limit, demonstrating progressive amplification of the \(\{11\bar{2}0\}\)-aligned state with increasing ring index.

Within the present modeling framework, the simulations reproduce the sign and radial amplification of the alignment metrics and are intended as a qualitative testbed rather than a parameter-calibrated quantitative fit to experiment. The absolute magnitudes of \(\kappa(r)\) depend on the chosen anisotropy strength, discretization, and noise level, and are therefore not expected to match experiment without parameter calibration.

The present results do not uniquely identify a single microscopic origin of the observed \(\{11\bar{2}0\}\) bias, but they do substantially narrow the range of viable explanations. In particular, the coexistence of opposite-polarity domains within the circular opening prior to the emergence of extended IDB traces rules out any interpretation based solely on single-domain initiation and vertex-driven geometric closure. More generally, the measured radial amplification of \(\kappa(r)\), the overall suppression of the \(\{1\bar{1}00\}\)-proximate tail, and the progressive narrowing of the orientation distribution together indicate that some form of propagation-mediated anisotropy is required. Within this broader class, several physically plausible realizations remain, including (i) step--terrace locking, in which boundary advance is biased through coupling to the underlying step structure;\cite{Xie-PRB-82-2749,Barbaray-DRM-8-314,Dimitrakopulos-PSSB-242-1617} (ii) faceting and coarsening driven by anisotropic interfacial energetics and growth kinetics, which can selectively stabilize and lengthen certain boundary segments;\cite{Hiramatsu-JCG-221-316} and (iii) chemically mediated anisotropic mobility or pinning,\cite{Roshko-JJAP-58-SC1050,Persson-SR-12-17987} in which local impurity incorporation or interfacial chemistry modifies propagation kinetics in an orientation-dependent manner. The present dataset is therefore best interpreted not as uniquely selecting one of these mechanisms, but as establishing a set of quantitative constraints that any successful mechanism must satisfy. Discriminating among these possibilities will require targeted follow-up measurements, for example through deliberate variation of substrate step structure, growth chemistry, or impurity conditions, combined with the same ring-resolved statistical analysis introduced here.

\section{Conclusions}
In summary, we establish a quantitative, statistics-based description of plan-view inversion domain boundary (IDB) orientations in polarity-mixed GaN lateral overgrowth within circular SiO\(_2\) mask openings. By combining polarity-resolved identification with automated IDB extraction and length-weighted orientation metrics, we show that long, straight IDB traces preferentially align with the \(\{11\bar{2}0\}\) family rather than the competing \(\{1\bar{1}00\}\) family. Crucially, this preference is observed in a mixed-polarity regime for which mask-boundary-imposed selection alone is insufficient: opposite-polarity domains already coexist within the openings prior to the emergence of extended IDB traces, indicating that the canonical single-domain, vertex-driven closure picture is not satisfied even though the macroscopic opening geometry remains circular.

The alignment strengthens systematically with propagation distance, as reflected in the increase of \(\kappa(r)\), the overall suppression of the \(\{1\bar{1}00\}\)-proximate tail, and the concurrent narrowing of the orientation distribution. Together, these trends indicate that the observed \(\{11\bar{2}0\}\)-biased alignment is progressively amplified during boundary propagation rather than being imposed solely by the circular opening geometry itself.

Minimal simulations provide a quantitative testbed for examining whether a minimal propagation-mediated anisotropy can reproduce the observed \(\{11\bar{2}0\}\) dominance and its radial amplification within the same circular-opening geometry. Although the present dataset does not uniquely identify a single microscopic origin, it places stringent quantitative constraints on any viable explanation. Rather than assigning the observed \(\{11\bar{2}0\}\) preference to a specific microscopic mechanism, our results show that some form of orientation-dependent propagation-mediated amplification is required in the present mixed-polarity regime. More broadly, the framework introduced here, which combines polarity-resolved mapping, reproducible orientation extraction, and distance-resolved order parameters, provides a transferable benchmark for disentangling geometric, kinetic, and chemical contributions to IDB alignment in GaN and related polar semiconductors.

\section{acknowledgments}
This work was supported by the National Research
Foundation of Korea (NRF) grant funded by the Korea
government (MSIT) (RS-2021-NR060087, RS-2023-
00240724) and through Korea Basic Science Institute
(National research Facilities and Equipment Center) grant
(2021R1A6C101A437) funded by the Ministry of Education.


\begin{thebibliography}{23}%
\makeatletter
\providecommand \@ifxundefined [1]{%
 \@ifx{#1\undefined}
}%
\providecommand \@ifnum [1]{%
 \ifnum #1\expandafter \@firstoftwo
 \else \expandafter \@secondoftwo
 \fi
}%
\providecommand \@ifx [1]{%
 \ifx #1\expandafter \@firstoftwo
 \else \expandafter \@secondoftwo
 \fi
}%
\providecommand \natexlab [1]{#1}%
\providecommand \enquote  [1]{``#1''}%
\providecommand \bibnamefont  [1]{#1}%
\providecommand \bibfnamefont [1]{#1}%
\providecommand \citenamefont [1]{#1}%
\providecommand \href@noop [0]{\@secondoftwo}%
\providecommand \href [0]{\begingroup \@sanitize@url \@href}%
\providecommand \@href[1]{\@@startlink{#1}\@@href}%
\providecommand \@@href[1]{\endgroup#1\@@endlink}%
\providecommand \@sanitize@url [0]{\catcode `\\12\catcode `\$12\catcode `\&12\catcode `\#12\catcode `\^12\catcode `\_12\catcode `\%12\relax}%
\providecommand \@@startlink[1]{}%
\providecommand \@@endlink[0]{}%
\providecommand \url  [0]{\begingroup\@sanitize@url \@url }%
\providecommand \@url [1]{\endgroup\@href {#1}{\urlprefix }}%
\providecommand \urlprefix  [0]{URL }%
\providecommand \Eprint [0]{\href }%
\providecommand \doibase [0]{http://dx.doi.org/}%
\providecommand \selectlanguage [0]{\@gobble}%
\providecommand \bibinfo  [0]{\@secondoftwo}%
\providecommand \bibfield  [0]{\@secondoftwo}%
\providecommand \translation [1]{[#1]}%
\providecommand \BibitemOpen [0]{}%
\providecommand \bibitemStop [0]{}%
\providecommand \bibitemNoStop [0]{.\EOS\space}%
\providecommand \EOS [0]{\spacefactor3000\relax}%
\providecommand \BibitemShut  [1]{\csname bibitem#1\endcsname}%
\let\auto@bib@innerbib\@empty
\bibitem [{\citenamefont {Hellman}(1998)}]{Hellman-IJNSR-3-11}%
  \BibitemOpen
  \bibfield  {author} {\bibinfo {author} {\bibfnamefont {E.}~\bibnamefont {Hellman}},\ }\href@noop {} {\bibfield  {journal} {\bibinfo  {journal} {Materials Research Society Internet Journal of Nitride Semiconductor Research}\ }\textbf {\bibinfo {volume} {3}},\ \bibinfo {pages} {11} (\bibinfo {year} {1998})}\BibitemShut {NoStop}%
\bibitem [{\citenamefont {Sumiya}\ and\ \citenamefont {Fuke}(2004)}]{Sumiya-IJNSR-9-1}%
  \BibitemOpen
  \bibfield  {author} {\bibinfo {author} {\bibfnamefont {M.}~\bibnamefont {Sumiya}}\ and\ \bibinfo {author} {\bibfnamefont {S.}~\bibnamefont {Fuke}},\ }\href@noop {} {\bibfield  {journal} {\bibinfo  {journal} {Internet Journal of Nitride Semiconductor Research}\ }\textbf {\bibinfo {volume} {9}},\ \bibinfo {pages} {1} (\bibinfo {year} {2004})}\BibitemShut {NoStop}%
\bibitem [{\citenamefont {Z{\'u}{\~n}iga-P{\'e}rez}\ \emph {et~al.}(2016)\citenamefont {Z{\'u}{\~n}iga-P{\'e}rez}, \citenamefont {Consonni}, \citenamefont {Lymperakis}, \citenamefont {Kong}, \citenamefont {Trampert}, \citenamefont {Fern{\'a}ndez-Garrido}, \citenamefont {Brandt}, \citenamefont {Renevier}, \citenamefont {Keller}, \citenamefont {Hestroffer}, \citenamefont {Wagner}, \citenamefont {Reparaz}, \citenamefont {Akyol}, \citenamefont {Rajan}, \citenamefont {Rennesson}, \citenamefont {Palacios},\ and\ \citenamefont {Feuillet}}]{Zuniga-APR-3-041303}%
  \BibitemOpen
  \bibfield  {author} {\bibinfo {author} {\bibfnamefont {J.}~\bibnamefont {Z{\'u}{\~n}iga-P{\'e}rez}}, \bibinfo {author} {\bibfnamefont {V.}~\bibnamefont {Consonni}}, \bibinfo {author} {\bibfnamefont {L.}~\bibnamefont {Lymperakis}}, \bibinfo {author} {\bibfnamefont {X.}~\bibnamefont {Kong}}, \bibinfo {author} {\bibfnamefont {A.}~\bibnamefont {Trampert}}, \bibinfo {author} {\bibfnamefont {S.}~\bibnamefont {Fern{\'a}ndez-Garrido}}, \bibinfo {author} {\bibfnamefont {O.}~\bibnamefont {Brandt}}, \bibinfo {author} {\bibfnamefont {H.}~\bibnamefont {Renevier}}, \bibinfo {author} {\bibfnamefont {S.}~\bibnamefont {Keller}}, \bibinfo {author} {\bibfnamefont {K.}~\bibnamefont {Hestroffer}}, \bibinfo {author} {\bibfnamefont {M.~R.}\ \bibnamefont {Wagner}}, \bibinfo {author} {\bibfnamefont {J.~S.}\ \bibnamefont {Reparaz}}, \bibinfo {author} {\bibfnamefont {F.}~\bibnamefont {Akyol}}, \bibinfo {author} {\bibfnamefont {S.}~\bibnamefont {Rajan}}, \bibinfo {author} {\bibfnamefont {S.}~\bibnamefont {Rennesson}}, \bibinfo
  {author} {\bibfnamefont {T.}~\bibnamefont {Palacios}}, \ and\ \bibinfo {author} {\bibfnamefont {G.}~\bibnamefont {Feuillet}},\ }\href {\doibase 10.1063/1.4963919} {\bibfield  {journal} {\bibinfo  {journal} {Appl. Phys. Rev.}\ }\textbf {\bibinfo {volume} {3}},\ \bibinfo {pages} {041303} (\bibinfo {year} {2016})}\BibitemShut {NoStop}%
\bibitem [{\citenamefont {Northrup}\ \emph {et~al.}(1996)\citenamefont {Northrup}, \citenamefont {Neugebauer},\ and\ \citenamefont {Romano}}]{Northrup-PRL-77-103}%
  \BibitemOpen
  \bibfield  {author} {\bibinfo {author} {\bibfnamefont {J.~E.}\ \bibnamefont {Northrup}}, \bibinfo {author} {\bibfnamefont {J.}~\bibnamefont {Neugebauer}}, \ and\ \bibinfo {author} {\bibfnamefont {L.~T.}\ \bibnamefont {Romano}},\ }\href@noop {} {\bibfield  {journal} {\bibinfo  {journal} {Phys. Rev. Lett.}\ }\textbf {\bibinfo {volume} {77}},\ \bibinfo {pages} {103} (\bibinfo {year} {1996})}\BibitemShut {NoStop}%
\bibitem [{\citenamefont {Umar}\ and\ \citenamefont {Sofo}(2021)}]{Umar-PRB-103-165305}%
  \BibitemOpen
  \bibfield  {author} {\bibinfo {author} {\bibfnamefont {M.~M.~F.}\ \bibnamefont {Umar}}\ and\ \bibinfo {author} {\bibfnamefont {J.~O.}\ \bibnamefont {Sofo}},\ }\href@noop {} {\bibfield  {journal} {\bibinfo  {journal} {Phys. Rev. B}\ }\textbf {\bibinfo {volume} {103}},\ \bibinfo {pages} {165305} (\bibinfo {year} {2021})}\BibitemShut {NoStop}%
\bibitem [{\citenamefont {Hwang}\ \emph {et~al.}(2024)\citenamefont {Hwang}, \citenamefont {Aigner}, \citenamefont {Metzger}, \citenamefont {Kummel},\ and\ \citenamefont {Cho}}]{Hwang-ACSAEM-6-3257}%
  \BibitemOpen
  \bibfield  {author} {\bibinfo {author} {\bibfnamefont {T.}~\bibnamefont {Hwang}}, \bibinfo {author} {\bibfnamefont {W.}~\bibnamefont {Aigner}}, \bibinfo {author} {\bibfnamefont {T.}~\bibnamefont {Metzger}}, \bibinfo {author} {\bibfnamefont {A.~C.}\ \bibnamefont {Kummel}}, \ and\ \bibinfo {author} {\bibfnamefont {K.}~\bibnamefont {Cho}},\ }\href@noop {} {\bibfield  {journal} {\bibinfo  {journal} {ACS Appl. Electron. Mater.}\ }\textbf {\bibinfo {volume} {6}},\ \bibinfo {pages} {3257} (\bibinfo {year} {2024})}\BibitemShut {NoStop}%
\bibitem [{\citenamefont {Beaumont}\ \emph {et~al.}(1998)\citenamefont {Beaumont}, \citenamefont {Gibart}, \citenamefont {Vaille}, \citenamefont {Haffouz}, \citenamefont {Nataf},\ and\ \citenamefont {Bouill{\'e}}}]{Beaumont-JCG-189-97}%
  \BibitemOpen
  \bibfield  {author} {\bibinfo {author} {\bibfnamefont {B.}~\bibnamefont {Beaumont}}, \bibinfo {author} {\bibfnamefont {P.}~\bibnamefont {Gibart}}, \bibinfo {author} {\bibfnamefont {M.}~\bibnamefont {Vaille}}, \bibinfo {author} {\bibfnamefont {S.}~\bibnamefont {Haffouz}}, \bibinfo {author} {\bibfnamefont {G.}~\bibnamefont {Nataf}}, \ and\ \bibinfo {author} {\bibfnamefont {A.}~\bibnamefont {Bouill{\'e}}},\ }\href@noop {} {\bibfield  {journal} {\bibinfo  {journal} {J. Cryst. Growth}\ }\textbf {\bibinfo {volume} {189}},\ \bibinfo {pages} {97} (\bibinfo {year} {1998})}\BibitemShut {NoStop}%
\bibitem [{\citenamefont {Hiramatsu}\ \emph {et~al.}(2000)\citenamefont {Hiramatsu}, \citenamefont {Nishiyama}, \citenamefont {Onishi}, \citenamefont {Mizutani}, \citenamefont {Narukawa}, \citenamefont {Motogaito}, \citenamefont {Miyake}, \citenamefont {Iyechika},\ and\ \citenamefont {Maeda}}]{Hiramatsu-JCG-221-316}%
  \BibitemOpen
  \bibfield  {author} {\bibinfo {author} {\bibfnamefont {K.}~\bibnamefont {Hiramatsu}}, \bibinfo {author} {\bibfnamefont {K.}~\bibnamefont {Nishiyama}}, \bibinfo {author} {\bibfnamefont {M.}~\bibnamefont {Onishi}}, \bibinfo {author} {\bibfnamefont {H.}~\bibnamefont {Mizutani}}, \bibinfo {author} {\bibfnamefont {M.}~\bibnamefont {Narukawa}}, \bibinfo {author} {\bibfnamefont {A.}~\bibnamefont {Motogaito}}, \bibinfo {author} {\bibfnamefont {H.}~\bibnamefont {Miyake}}, \bibinfo {author} {\bibfnamefont {Y.}~\bibnamefont {Iyechika}}, \ and\ \bibinfo {author} {\bibfnamefont {T.}~\bibnamefont {Maeda}},\ }\href@noop {} {\bibfield  {journal} {\bibinfo  {journal} {J. Cryst. Growth}\ }\textbf {\bibinfo {volume} {221}},\ \bibinfo {pages} {316} (\bibinfo {year} {2000})}\BibitemShut {NoStop}%
\bibitem [{\citenamefont {Kim}\ \emph {et~al.}(2018)\citenamefont {Kim}, \citenamefont {Lee}, \citenamefont {Jang}, \citenamefont {Kim},\ and\ \citenamefont {Kim}}]{Kim-JAC-51-1551}%
  \BibitemOpen
  \bibfield  {author} {\bibinfo {author} {\bibfnamefont {H.~S.}\ \bibnamefont {Kim}}, \bibinfo {author} {\bibfnamefont {H.}~\bibnamefont {Lee}}, \bibinfo {author} {\bibfnamefont {D.}~\bibnamefont {Jang}}, \bibinfo {author} {\bibfnamefont {D.}~\bibnamefont {Kim}}, \ and\ \bibinfo {author} {\bibfnamefont {C.}~\bibnamefont {Kim}},\ }\href@noop {} {\bibfield  {journal} {\bibinfo  {journal} {J. Appl. Crystallography}\ }\textbf {\bibinfo {volume} {51}},\ \bibinfo {pages} {1551} (\bibinfo {year} {2018})}\BibitemShut {NoStop}%
\bibitem [{\citenamefont {Lee}\ \emph {et~al.}(2019)\citenamefont {Lee}, \citenamefont {Jang}, \citenamefont {Kim},\ and\ \citenamefont {Kim}}]{Lee-JAC-52-532}%
  \BibitemOpen
  \bibfield  {author} {\bibinfo {author} {\bibfnamefont {H.}~\bibnamefont {Lee}}, \bibinfo {author} {\bibfnamefont {D.}~\bibnamefont {Jang}}, \bibinfo {author} {\bibfnamefont {D.}~\bibnamefont {Kim}}, \ and\ \bibinfo {author} {\bibfnamefont {C.}~\bibnamefont {Kim}},\ }\href@noop {} {\bibfield  {journal} {\bibinfo  {journal} {J. Appl. Crystallography}\ }\textbf {\bibinfo {volume} {52}},\ \bibinfo {pages} {532} (\bibinfo {year} {2019})}\BibitemShut {NoStop}%
\bibitem [{\citenamefont {Zhang}\ \emph {et~al.}(2023)\citenamefont {Zhang}, \citenamefont {Persson}, \citenamefont {Chen}, \citenamefont {Papamichail}, \citenamefont {Tran}, \citenamefont {Persson}, \citenamefont {Paskov},\ and\ \citenamefont {Darakchieva}}]{Zhang-CGD-23-1049}%
  \BibitemOpen
  \bibfield  {author} {\bibinfo {author} {\bibfnamefont {H.}~\bibnamefont {Zhang}}, \bibinfo {author} {\bibfnamefont {I.}~\bibnamefont {Persson}}, \bibinfo {author} {\bibfnamefont {J.-T.}\ \bibnamefont {Chen}}, \bibinfo {author} {\bibfnamefont {A.}~\bibnamefont {Papamichail}}, \bibinfo {author} {\bibfnamefont {D.~Q.}\ \bibnamefont {Tran}}, \bibinfo {author} {\bibfnamefont {P.~O.}\ \bibnamefont {Persson}}, \bibinfo {author} {\bibfnamefont {P.~P.}\ \bibnamefont {Paskov}}, \ and\ \bibinfo {author} {\bibfnamefont {V.}~\bibnamefont {Darakchieva}},\ }\href@noop {} {\bibfield  {journal} {\bibinfo  {journal} {Crystal Growth \& Design}\ }\textbf {\bibinfo {volume} {23}},\ \bibinfo {pages} {1049} (\bibinfo {year} {2023})}\BibitemShut {NoStop}%
\bibitem [{\citenamefont {Li}\ \emph {et~al.}(2001)\citenamefont {Li}, \citenamefont {Sumiya}, \citenamefont {Fuke}, \citenamefont {Yang}, \citenamefont {Que}, \citenamefont {Suzuki},\ and\ \citenamefont {Fukuda}}]{Li-JAP-90-4219}%
  \BibitemOpen
  \bibfield  {author} {\bibinfo {author} {\bibfnamefont {D.}~\bibnamefont {Li}}, \bibinfo {author} {\bibfnamefont {M.}~\bibnamefont {Sumiya}}, \bibinfo {author} {\bibfnamefont {S.}~\bibnamefont {Fuke}}, \bibinfo {author} {\bibfnamefont {D.}~\bibnamefont {Yang}}, \bibinfo {author} {\bibfnamefont {D.}~\bibnamefont {Que}}, \bibinfo {author} {\bibfnamefont {Y.}~\bibnamefont {Suzuki}}, \ and\ \bibinfo {author} {\bibfnamefont {Y.}~\bibnamefont {Fukuda}},\ }\href@noop {} {\bibfield  {journal} {\bibinfo  {journal} {J. Appl. Phys.}\ }\textbf {\bibinfo {volume} {90}},\ \bibinfo {pages} {4219} (\bibinfo {year} {2001})}\BibitemShut {NoStop}%
\bibitem [{\citenamefont {Zhuang}\ and\ \citenamefont {Edgar}(2005)}]{Zhuang-MSER-48-1}%
  \BibitemOpen
  \bibfield  {author} {\bibinfo {author} {\bibfnamefont {D.}~\bibnamefont {Zhuang}}\ and\ \bibinfo {author} {\bibfnamefont {J.}~\bibnamefont {Edgar}},\ }\href@noop {} {\bibfield  {journal} {\bibinfo  {journal} {Mater. Sci. Eng. R}\ }\textbf {\bibinfo {volume} {48}},\ \bibinfo {pages} {1} (\bibinfo {year} {2005})}\BibitemShut {NoStop}%
\bibitem [{\citenamefont {Schindelin}\ \emph {et~al.}(2012)\citenamefont {Schindelin}, \citenamefont {Arganda-Carreras}, \citenamefont {Frise}, \citenamefont {Kaynig}, \citenamefont {Longair}, \citenamefont {Pietzsch}, \citenamefont {Preibisch}, \citenamefont {Rueden}, \citenamefont {Saalfeld}, \citenamefont {Schmid} \emph {et~al.}}]{Schindelin-NM-9-676}%
  \BibitemOpen
  \bibfield  {author} {\bibinfo {author} {\bibfnamefont {J.}~\bibnamefont {Schindelin}}, \bibinfo {author} {\bibfnamefont {I.}~\bibnamefont {Arganda-Carreras}}, \bibinfo {author} {\bibfnamefont {E.}~\bibnamefont {Frise}}, \bibinfo {author} {\bibfnamefont {V.}~\bibnamefont {Kaynig}}, \bibinfo {author} {\bibfnamefont {M.}~\bibnamefont {Longair}}, \bibinfo {author} {\bibfnamefont {T.}~\bibnamefont {Pietzsch}}, \bibinfo {author} {\bibfnamefont {S.}~\bibnamefont {Preibisch}}, \bibinfo {author} {\bibfnamefont {C.}~\bibnamefont {Rueden}}, \bibinfo {author} {\bibfnamefont {S.}~\bibnamefont {Saalfeld}}, \bibinfo {author} {\bibfnamefont {B.}~\bibnamefont {Schmid}},  \emph {et~al.},\ }\href@noop {} {\bibfield  {journal} {\bibinfo  {journal} {Nature Methods}\ }\textbf {\bibinfo {volume} {9}},\ \bibinfo {pages} {676} (\bibinfo {year} {2012})}\BibitemShut {NoStop}%
\bibitem [{\citenamefont {Arganda-Carreras}\ \emph {et~al.}(2017)\citenamefont {Arganda-Carreras}, \citenamefont {Kaynig}, \citenamefont {Rueden}, \citenamefont {Eliceiri}, \citenamefont {Schindelin}, \citenamefont {Cardona},\ and\ \citenamefont {Sebastian~Seung}}]{Arganda-BI-33-2424}%
  \BibitemOpen
  \bibfield  {author} {\bibinfo {author} {\bibfnamefont {I.}~\bibnamefont {Arganda-Carreras}}, \bibinfo {author} {\bibfnamefont {V.}~\bibnamefont {Kaynig}}, \bibinfo {author} {\bibfnamefont {C.}~\bibnamefont {Rueden}}, \bibinfo {author} {\bibfnamefont {K.~W.}\ \bibnamefont {Eliceiri}}, \bibinfo {author} {\bibfnamefont {J.}~\bibnamefont {Schindelin}}, \bibinfo {author} {\bibfnamefont {A.}~\bibnamefont {Cardona}}, \ and\ \bibinfo {author} {\bibfnamefont {H.}~\bibnamefont {Sebastian~Seung}},\ }\href@noop {} {\bibfield  {journal} {\bibinfo  {journal} {Bioinformatics}\ }\textbf {\bibinfo {volume} {33}},\ \bibinfo {pages} {2424} (\bibinfo {year} {2017})}\BibitemShut {NoStop}%
\bibitem [{\citenamefont {Jolliffe}\ and\ \citenamefont {Cadima}(2016)}]{Jolliffe-PTRSA-374-20150202}%
  \BibitemOpen
  \bibfield  {author} {\bibinfo {author} {\bibfnamefont {I.~T.}\ \bibnamefont {Jolliffe}}\ and\ \bibinfo {author} {\bibfnamefont {J.}~\bibnamefont {Cadima}},\ }\href {\doibase 10.1098/rsta.2015.0202} {\bibfield  {journal} {\bibinfo  {journal} {Philosophical Transactions of the Royal Society A: Mathematical, Physical and Engineering Sciences}\ }\textbf {\bibinfo {volume} {374}},\ \bibinfo {pages} {20150202} (\bibinfo {year} {2016})}\BibitemShut {NoStop}%
\bibitem [{\citenamefont {Osher}\ and\ \citenamefont {Fedkiw}(2003)}]{Osher-LSM}%
  \BibitemOpen
  \bibfield  {author} {\bibinfo {author} {\bibfnamefont {S.}~\bibnamefont {Osher}}\ and\ \bibinfo {author} {\bibfnamefont {R.~P.}\ \bibnamefont {Fedkiw}},\ }\href@noop {} {\emph {\bibinfo {title} {Level set methods and dynamic implicit surfaces}}}\ (\bibinfo  {publisher} {Springer, New York},\ \bibinfo {year} {2003})\BibitemShut {NoStop}%
\bibitem [{\citenamefont {Du}\ \emph {et~al.}(2005)\citenamefont {Du}, \citenamefont {Srolovitz}, \citenamefont {Coltrin},\ and\ \citenamefont {Mitchell}}]{Du-PRL-95-155503}%
  \BibitemOpen
  \bibfield  {author} {\bibinfo {author} {\bibfnamefont {D.}~\bibnamefont {Du}}, \bibinfo {author} {\bibfnamefont {D.~J.}\ \bibnamefont {Srolovitz}}, \bibinfo {author} {\bibfnamefont {M.~E.}\ \bibnamefont {Coltrin}}, \ and\ \bibinfo {author} {\bibfnamefont {C.~C.}\ \bibnamefont {Mitchell}},\ }\href@noop {} {\bibfield  {journal} {\bibinfo  {journal} {Phys. Rev. Lett.}\ }\textbf {\bibinfo {volume} {95}},\ \bibinfo {pages} {155503} (\bibinfo {year} {2005})}\BibitemShut {NoStop}%
\bibitem [{\citenamefont {Xie}\ \emph {et~al.}(1999)\citenamefont {Xie}, \citenamefont {Seutter}, \citenamefont {Zhu}, \citenamefont {Zheng}, \citenamefont {Wu},\ and\ \citenamefont {Tong}}]{Xie-PRB-82-2749}%
  \BibitemOpen
  \bibfield  {author} {\bibinfo {author} {\bibfnamefont {M.}~\bibnamefont {Xie}}, \bibinfo {author} {\bibfnamefont {S.}~\bibnamefont {Seutter}}, \bibinfo {author} {\bibfnamefont {W.}~\bibnamefont {Zhu}}, \bibinfo {author} {\bibfnamefont {L.}~\bibnamefont {Zheng}}, \bibinfo {author} {\bibfnamefont {H.}~\bibnamefont {Wu}}, \ and\ \bibinfo {author} {\bibfnamefont {S.}~\bibnamefont {Tong}},\ }\href@noop {} {\bibfield  {journal} {\bibinfo  {journal} {Phys. Rev. Lett.}\ }\textbf {\bibinfo {volume} {82}},\ \bibinfo {pages} {2749} (\bibinfo {year} {1999})}\BibitemShut {NoStop}%
\bibitem [{\citenamefont {Barbaray}\ \emph {et~al.}(1999)\citenamefont {Barbaray}, \citenamefont {Potin}, \citenamefont {Ruterana},\ and\ \citenamefont {Nouet}}]{Barbaray-DRM-8-314}%
  \BibitemOpen
  \bibfield  {author} {\bibinfo {author} {\bibfnamefont {B.}~\bibnamefont {Barbaray}}, \bibinfo {author} {\bibfnamefont {V.}~\bibnamefont {Potin}}, \bibinfo {author} {\bibfnamefont {P.}~\bibnamefont {Ruterana}}, \ and\ \bibinfo {author} {\bibfnamefont {G.}~\bibnamefont {Nouet}},\ }\href@noop {} {\bibfield  {journal} {\bibinfo  {journal} {Diam. Relat. Mater.}\ }\textbf {\bibinfo {volume} {8}},\ \bibinfo {pages} {314} (\bibinfo {year} {1999})}\BibitemShut {NoStop}%
\bibitem [{\citenamefont {Dimitrakopulos}\ \emph {et~al.}(2005)\citenamefont {Dimitrakopulos}, \citenamefont {Sanchez}, \citenamefont {Komninou}, \citenamefont {Kehagias}, \citenamefont {Karakostas}, \citenamefont {Nouet},\ and\ \citenamefont {Ruterana}}]{Dimitrakopulos-PSSB-242-1617}%
  \BibitemOpen
  \bibfield  {author} {\bibinfo {author} {\bibfnamefont {G.}~\bibnamefont {Dimitrakopulos}}, \bibinfo {author} {\bibfnamefont {A.}~\bibnamefont {Sanchez}}, \bibinfo {author} {\bibfnamefont {P.}~\bibnamefont {Komninou}}, \bibinfo {author} {\bibfnamefont {T.}~\bibnamefont {Kehagias}}, \bibinfo {author} {\bibfnamefont {T.}~\bibnamefont {Karakostas}}, \bibinfo {author} {\bibfnamefont {G.}~\bibnamefont {Nouet}}, \ and\ \bibinfo {author} {\bibfnamefont {P.}~\bibnamefont {Ruterana}},\ }\href@noop {} {\bibfield  {journal} {\bibinfo  {journal} {Physica Status Solidi (b)}\ }\textbf {\bibinfo {volume} {242}},\ \bibinfo {pages} {1617} (\bibinfo {year} {2005})}\BibitemShut {NoStop}%
\bibitem [{\citenamefont {Roshko}\ \emph {et~al.}(2019)\citenamefont {Roshko}, \citenamefont {Brubaker}, \citenamefont {Blanchard}, \citenamefont {Harvey},\ and\ \citenamefont {Bertness}}]{Roshko-JJAP-58-SC1050}%
  \BibitemOpen
  \bibfield  {author} {\bibinfo {author} {\bibfnamefont {A.}~\bibnamefont {Roshko}}, \bibinfo {author} {\bibfnamefont {M.}~\bibnamefont {Brubaker}}, \bibinfo {author} {\bibfnamefont {P.}~\bibnamefont {Blanchard}}, \bibinfo {author} {\bibfnamefont {T.}~\bibnamefont {Harvey}}, \ and\ \bibinfo {author} {\bibfnamefont {K.}~\bibnamefont {Bertness}},\ }\href@noop {} {\bibfield  {journal} {\bibinfo  {journal} {Jpn. J. Appl. Phys.}\ }\textbf {\bibinfo {volume} {58}},\ \bibinfo {pages} {SC1050} (\bibinfo {year} {2019})}\BibitemShut {NoStop}%
\bibitem [{\citenamefont {Persson}\ \emph {et~al.}(2022)\citenamefont {Persson}, \citenamefont {Papamichail}, \citenamefont {Darakchieva},\ and\ \citenamefont {Persson}}]{Persson-SR-12-17987}%
  \BibitemOpen
  \bibfield  {author} {\bibinfo {author} {\bibfnamefont {A.~R.}\ \bibnamefont {Persson}}, \bibinfo {author} {\bibfnamefont {A.}~\bibnamefont {Papamichail}}, \bibinfo {author} {\bibfnamefont {V.}~\bibnamefont {Darakchieva}}, \ and\ \bibinfo {author} {\bibfnamefont {P.~O.}\ \bibnamefont {Persson}},\ }\href@noop {} {\bibfield  {journal} {\bibinfo  {journal} {Sci. Rep.}\ }\textbf {\bibinfo {volume} {12}},\ \bibinfo {pages} {17987} (\bibinfo {year} {2022})}\BibitemShut {NoStop}%
\end{thebibliography}

%

\end{document}